\documentstyle[aps,prl,psfig,twocolumn]{revtex}
\begin{document}
\wideabs{
\draft \title{Microscopic Theory of Heterogeneity and  Non-Exponential Relaxations in Supercooled Liquids}
\author{Xiaoyu Xia and Peter G. Wolynes\footnotemark[1]}
\address{Department of Physics and Chemistry, University of Illinois, Urbana, IL, 61801\\}
\address{\footnotemark[1] Present address: Department of Chemistry and Biochemistry, University of California, San Diego, 9500 Gilman Drive, La Jolla, CA 92093}

\maketitle

\begin{abstract}
{Recent experiments and computer simulations show that supercooled liquids around the glass transition temperature are ``dynamically heterogeneous" \cite{Sillescu99}. Such heterogeneity
is expected from the random first order transition theory of the glass transition.  Using a microscopic approach based on this theory, we derive a relation
between the departure from Debye relaxation as characterized by the $\beta$ value of a stretched exponential response function $\phi(t) =e^{-(t/ \tau _{KWW})^{\beta}}$, and the fragility of the liquid.  The $\beta$ value is also predicted to depend on temperature and to vanish as the ideal glass transition is approached at
the Kauzmann temperature.}
\end{abstract}

}

The striking universality of relaxation dynamics in supercooled liquids has remained intriguing for decades.  In addition to the overall dramatic slowing of transport as the glass transition is approached, one finds the
emergence of a strongly non-exponential  approach to equilibrium when a supercooled liquid is perturbed.  This contrasts with the behavior of chemically
simple liquids at higher temperatures, where only a single time scale for a highly 
exponential structural relaxation is usually encountered for times beyond the
vibrational time scales.  Both the range of time scales and the median magnitude of the relaxation time require explanation.

Several theoretical threads lead to the notion that the universal behavior of supercooled liquids arises from proximity to an underlying random first order
transition \cite{KW87a,KT87,KW87b,KTW89,MP99} which is found in mean field theories of spin glass without reflection symmetry \cite{GKS85,Gardner85,MPV87}, and in mode coupling \cite{BGS84,Gotze91} and density functional \cite{SW84,SSW85,DV99} approaches to the structural glass transition.  This picture explains both the breakdown of
simple collisional theories of transport that apply to high temperature liquids at a characteristic temperature $T_A$ and the impending entropy crisis of supercooled liquids first discovered
by Simon and brilliantly emphasized by Kauzmann \cite{Kauz43} at the temperature $T_K$.  Furthermore the scenario suggests that the finite range of the underlying force modifies mean field
behavior between $T_A$ and $T_K$ in a way that leads to an intricate ``mosaic" structure of a glassy
fluid in which mesoscopic local regions are individually each in an aperiodic minimum
but are separated by more mobile domain walls which are strained and quite far from local minima structures \cite{Wolynes89}.  Relaxation of the elements of the mosaic, reconfiguring to other low energy structures, leads to the slow relaxation and a scaling treatment of the median relaxation time yields the venerable Vogel-Fulcher law, $\tau =\tau_0 e^{\frac{DT_0}{T-T_0}}$ \cite{KTW89}. $D$, called the fragility, determines the apparent
size of deviations from an Arrhenius law.  Most recently a microscopic calculation of the coefficient $D$ has yielded good agreement with experiments
for a range of liquids \cite{XW00}.  In this paper, we address the predictions of the
mosaic picture for the
dispersion of relaxation times.

Relaxation in supercooled liquids is well approximated by the stretched exponential or Kohlrausch-Williams-Watts (KWW) formula $\phi(t)=e^{-(t/ \tau _{KWW})^{\beta}}$.  A study conducted by B\"{o}hmer et al. \cite{BNAP93} on over sixty glass formers
around $T_g$ shows $\beta$ and $D$ are strongly correlated.  The smallest $\beta$ is found for the most fragile liquids.  The heterogeneity of time scales
suggests possible heterogeneity in space.

The spatial heterogeneity implied by the mosaic picture has received strong
support from recent computer simulations \cite{KDPPG97,DDKPPG98} and laboratory experiments, especially direct
measurements using 4-D NMR \cite{TWHFSS98} and optical hole burning \cite{CE95} techniques.
In the mosaic picture, different regions of the 
supercooled liquid will relax in different ways depending on how stable the local structure is, but for time scales much
longer than the median, the system will behave homogeneously since the neighboring elements of a mosaic cell will likely also have reconfigured.  In this picture, despite the presence of some dynamical averaging, 
the response function can be viewed to a good approximation as arising from a
relaxation time distribution $P(\tau )$
\begin{equation}\label{eq:step}
\phi(t)=\int e^{-t/\tau } P(\tau)d\tau.
\end{equation}
This will resemble a KWW formula with the $\beta$ parameter determined by the explicit form
of $P(\tau )$, largely by the breadth of the distribution.

According to the random first order transition theory \cite{KW87a,KW87b,KTW89,XW00} in supercooled liquids the relaxation of an individual mosaic element is an activated process.  
The driving force for any local region to escape from
one of the metastable states predicted by a meanfield free energy functional is the configurational entropy of the other states to which it might hop.  Working against this is a cost due to surface energy since the domain wall is not in a low free energy configuration.  
Creating a droplet costs free energy that depends on the radius 
of the droplet.  Much as in conventional nucleation, one finds
\begin{equation}\label{eq:mffe}
F(r)=-\frac{4}{3}\pi Ts_cr^3+4 \pi \sigma (r) r^2.
\end{equation}
Here $s_c$ is the configurational entropy density which drives random first order transition.  A novel feature is that the multiplicity of states leads to a renormalization of surface tension, $\sigma(r)=\sigma _0(\frac{r_0}{r})^{1/2}$ 
\cite{XW00}, where $r_0$ is the interparticle spacing.  The typical free energy barrier is determined by the maximum of Eq.(\ref{eq:mffe}) as a function of $r$, giving
\begin{equation}\label{eq:fnbr}
\Delta F^{\ddagger}=\frac{3\pi\sigma^2_0r_0}{ Ts_c}=\frac{3\pi\sigma^2_0r_0}{T\Delta \tilde{c_p}} \frac{T_K}{T-
T_K}=k_BTD\frac{T_K}{T-T_K},
\end{equation}
since $s_c=\Delta \tilde{c_p}(T)\frac{T-T_K}{T_K}$, where $\Delta \tilde{c_p}(T)$ is the specific heat jump per unit volume at the transition.   
The microscopic theory of $\sigma_0$ gives \cite{XW00} 
\begin{equation}\label{eq:dval}
D=\frac{27}{16}\pi \frac{nk_B}{\Delta \tilde{c_p}}\ln ^2\frac{\alpha _Lr_0^2}{\pi e},
\end{equation}
where $\alpha_L$ is the square inverse of the Lindemann ratio of the maximum
vibrational displacement around an aperiodic minimum which is globally stable
$\alpha_Lr_0^2=100$.  The resulting formula for the fragility, $D=32R/\Delta c_p$ fits a wide range of substances quite well.  ($\Delta c_p$ is heat capacity jump per mole and $R$ is the gas constant.)  The typical droplet size in the mosaic corresponding
to the relaxation barrier turns out to be
\begin{equation}\label{eq:col}
r^{\ddagger}=(\frac{2}{3\pi \ln \frac{\alpha_Lr_0^2}{\pi e}})^{2/3}(\frac{DT_K}{T-
T_K})^{2/3}r_0.
\end{equation}
At $T_g$, this formula for $r^{\ddagger}$ gives a correlation length about 5 molecular radii 
(or a few nanometers) \cite{XW00}, which is consistent with the experimental findings
\cite{TWHFSS98} but much larger than the regions envisioned in the Adam-Gibbs approach \cite{AG65}.  The configuration of the supercooled liquid is separated into domains with average size 
$\overline{r^*}\approx 1.6r^{\ddagger}$ which signals where the free energy Eq. (2) vanishes.  According to the microscopic theory both $r^{\ddagger}$ and $\overline{r^*}$ (in the unit of molecular distance) are nearly universal
functions of the relaxation time at which the response occurs.  This is because
the Lindemann ratio is nearly universal for a wide range of substances.  Each domain corresponds to a minimum of the free energy functional,
but these domains would vary somewhat in size since there are many different local 
minimum states.  They will be separated by thin mobile sections.  The fluctuation of energy of each state can also be said to reflect the idea that the configurational entropy (at fixed energy!) itself fluctuates according to the usual Landau formula \cite{Landau69}
\begin{equation}\label{eq:sflu}
\delta s_c=\sqrt{k_B\Delta \tilde{c_p}/V^{\ddagger}}, 
\end{equation}
where $V^{\ddagger}=\frac{4}{3}\pi r^{\ddagger 3}$ is the volume of the average
domain.  The fluctuation in configuration entropy, the driving force, results
in a corresponding variation in free energy barriers for each mosaic element and therefore gives a distribution of relaxation times.

\begin{figure}
\psfig{file=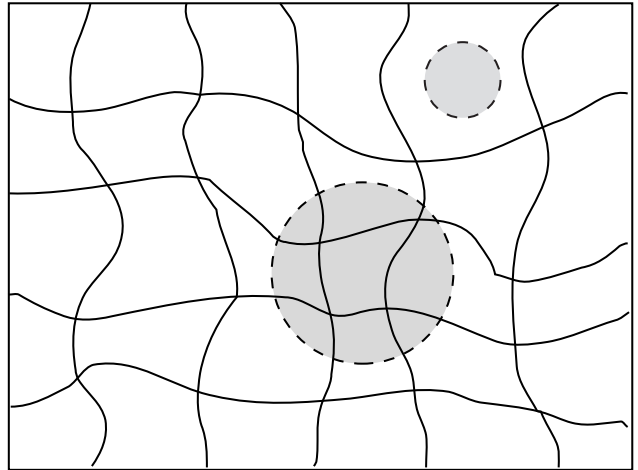,width=3.3in}
\caption
{An illustration of the ``mosaic structure" of supercooled liquids.  The mosaic
pieces are not necessarily of the same size due to the fluctuations in the driving force, configurational entropy.  The system escapes from a local metastable configuration by an activated process equivalent to forming a liquid-like droplet inside a mosaic element.  For
droplets with size much smaller than the mosaic (as shown with the small circle), the
droplet shape and its surface energy cost are well described by the infinite system result  Eq.(2).  For transition state droplets that would seem larger than the typical mosaic elements (as shown with the large circle),
the surface energy cost will be much smaller than $\sigma r^2$ since the formation of the such droplet will break through boundaries of the preexisting domain.  In this case, the free energy barrier to form such a large droplet will be close to $\Delta F_0$, the most probable barrier determined by the macroscopic configurational entropy density. 
}
\end{figure}

\begin{figure}
\psfig{file=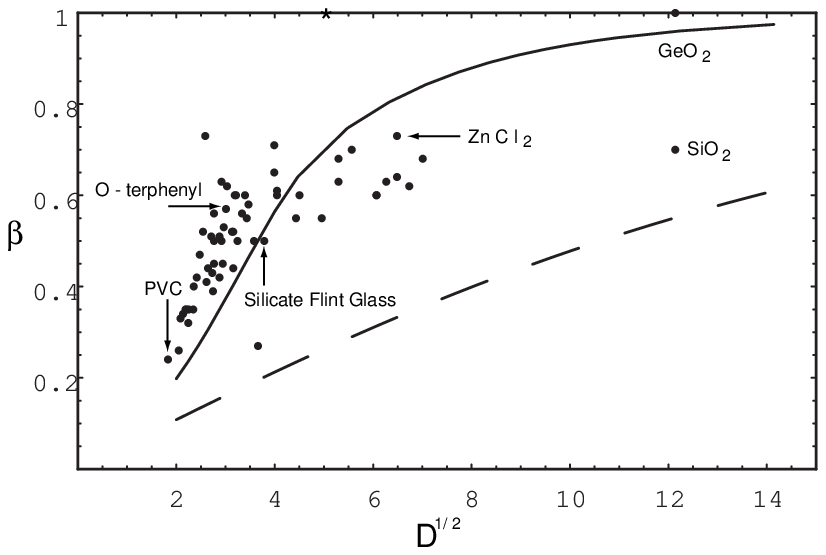,width=3.3in}
\caption
{Correlation between the $\beta$ value for non-Debye relaxation and square root of fragility $D$.  The points are from a wide range of experiments collected in [18].  It is important to recognize that in this sort of ``meta-analysis" of experiments that different investigators use somewhat different
ways of fitting relaxation data and defining the glass transition
time scale in the laboratory, leading to a spread of values for the measured
$\beta$ as reported by B\"{o}hmer et al. [18].  For example the $\beta$ value for n-propanol ($*$) is actually measured for a glass transition with a time scale of $10^{-2}$ seconds [28], 4 orders of magnitude smaller than the usual glass transition relaxation time, 100 seconds, leading to a larger $\beta$.  The pure $\mathrm{SiO_2}$ number seems anomalous and deserves careful remeasurement since it differs considerably from the $\mathrm{GeO_2}$ result.  The dashed line is obtained assuming a Gaussian free energy barrier distribution.  It is quite linear.  This approximation tends to overestimate the width of the distribution resulting in a smaller $\beta$.  Results from the more realistic cutoff distribution
taking into account the constraints imposed by the mosaic structures yield the solid line.  Neither theoretical result contains any fitting parameters, owing to the universal character of the surface energy costs, obtained by microscopic estimates of the domain size of $T_g$.
}
\end{figure}

To find the $\beta$ value, we must construct an explicit distribution of free energy barriers based on such
dynamically fluctuating mosaic structure.  To set the stage, let us first see what happens if we assume the
distribution of relaxation times to be Gaussian resulting from small fluctuations.  
This calculation parallels one carried out by Ediger to infer the domain size
in supercooled liquids \cite{Ediger98}.   
This approximation already yields a qualitatively (but not quantitatively!) correct correlation of $\beta$ with $D$.  The time correlation function can be rewritten as 
\begin{equation}\label{eq:step2}
\phi(t)=\int e^{-t/\tau(\Delta F)} P(\Delta F)d\Delta F.
\end{equation}
Here $\tau(\Delta F)=\tau_0 e^{\Delta F/k_BT}$.  If $P(\Delta F)$ is Gausssian, the relaxation function is not precisely a stretched exponential but is well
fitted by one with a $\beta$ value given by
\begin{equation}\label{eq:beta1}
\beta =[1+(\delta \Delta F/k_BT)^2]^{-\frac{1}{2}}, 
\end{equation}
where $\delta \Delta F$ is the width of the Gaussian distribution 
$P(\Delta F)=\frac{1}{\sqrt{2\pi \delta \Delta F^2}}e^{-(\Delta F-\Delta F_0)^2/2\delta \Delta F^2}$.  Using Eq.(3)-(6), we find $\frac{\delta \Delta F}{\Delta F_0}\approx \frac{\delta s_c}{<s_c>}=\frac{1}{2\sqrt{D}}$.  Therefore we have
\begin{equation}\label{eq:beta2}
\beta =[1+(\frac{\Delta F_0(T)}{2k_BT\sqrt{D}})^2]^{-\frac{1}{2}}. 
\end{equation}
This formula should be valid for a range of substances again because of the universal nature of the correlation volume at a given time scale predicted
by the microscopic theory of fragility \cite{XW00}.
The typical barrier height at $T_g$ as conventionally defined with a relaxation time of $10^2$ seconds corresponds to $\Delta F_0\approx 37 k_BT_g$ thus this estimate gives for $\beta$ at the laboratory glass transition temperature
$\beta_G \approx \frac{\sqrt{D}}{18.5}$.  
The result is also shown on Fig.(2) as a dashed line.  The data in the graph
are the $\beta$ values measured at $T_g$ for a wide variety of substances.  We see that this formula suggests correctly that more fragile liquids have smaller $\beta$ values near their glass transition.  Also we see that as the temperature
is lowered, the most probable barrier height $\Delta F_0(T)$ increases rapidly, giving a smaller value of $\beta$ ultimately vanishing at $T_K$ if
available time permitted measurement and equilibration at such low temperature.  Such a temperature dependence was found in the detailed experimental study for o-terphenyl by Dixon and Nagel \cite{DN88}.  While it seems experiment agrees that $\beta$ approaches zero around $T_K$ consistent with Eq.(9),
some theories do not give such a relation.  Free volume theory gives a lower limit of
$\beta=\frac{2}{3}$ for example in \cite{CG81}, a value that has been surpassed in experiments \cite{DN88}.

Although Eq.(9) gives the right trend of $\beta -D$ correlation, it is not quantitatively accurate.  A more careful analysis of the implications of the
mosaic structure is needed.  The mosaic structure of the random first order transition theory implies the existence of 
large correction to the Gaussian result since the fluctuations in $\Delta F$ scale in the same way as
$\Delta F_0$.  They are thus of a similar magnitude since $r^{\ddagger}$ is
of the same size as $r^*$.  We shall now show that 
a reasonably realistic distribution easily comes out from a simple
model of the dynamic mosaic structure of supercooled liquids.  First, it is clear that even if the distribution of configurational entropy were precisely Gaussian,
the free energy barrier distribution would not be.  We should use $P(\Delta F)d\Delta F=P(s_c)ds_c$ to get the precise distribution.  Second, domains will not all be of equal size,  instead, there is a distribution of size determined again by fluctuations in configurational
entropy as we have $r^{\ddag}\sim (\frac{1}{s_c})^{1.5}$.  Most important, when a certain domain with small size $r'$ already exists, the free energy barrier for overturning that domain will be smaller than what Eq.(3) predicts since this equation assumes spherical transition state droplets.  This activated droplet will be modified because of the pre-existing boundaries.  When the fluctuating size within the spherical droplet is comparable or bigger than $r'$ these boundary effects
will limit the size of the barrier.  This effect may be roughly described by simply assuming there is a ``cutoff"
in the free energy barrier distribution near the most probable one.  Similarly if a neighboring region has already flipped this will make it easier to reconfigure the domain under consideration.  Both of these effects suggest the real distribution
in the mosaic can be better approximated in a piece-wise manner, 
\[ P(\Delta F) = \left\{ \begin{array}{ll}
P_f(\Delta F) & \mbox{$\Delta F \leq \Delta F_0$} \\
C \delta (\Delta F-\Delta F_0) & \mbox{$\Delta F > \Delta F_0$}
\end{array}
\right. \]
where $C$ is a normalization coefficient so $\int_{-\infty}^{+\infty} P(\Delta F)d\Delta F=1$ and $P_f(\Delta F)$ is a function determined by 
$P_f(\Delta F)d\Delta F=P(s_c)ds_c$.  Here $\Delta F_0$ is the barrier assuming
no fluctuation in configurational entropy.
This cutoff distribution of free energy barriers still yields a $\phi (t)$ that fits the KWW formula.  The resulting $\beta$ value is, however, different.  The results from the cutoff distribution for $\beta$ at $T_g$ are shown in Fig.(2) as solid line.  The slope of $\beta$ versus $\sqrt{D}$ is considerably increased and $\beta$ now saturates to 1 for liquids with $D>150$, a value characteristic of the so-called strong liquids.  The improvement of agreement over the pure Gaussian is significant, buttressing the case for a dynamic mosaic structure.

Above $T_A$, the critical temperature predicted by mode coupling theory for
dynamic slowing, relaxation is dominated by collision between molecules.  Here dynamics 
can be described by the Debye law with a single exponential.  We see that between $T_A$ and $T_K$, the deviation from the
Debye law will become more significant upon cooling as a consequence of increasing heterogeneity just as in the Gaussian analysis.  For Gaussian distribution, one finds $\beta \approx \frac{1}{\sqrt{D}} \frac{T-T_K}{T_K}$ at temperature close to $T_g$.  In Fig.(3), we also plot the temperature dependence of $\beta$ predicted by the more accurate cutoff distribution for o-terphenyl and compare it with experiments \cite{DN88}.  In computing the $T$ dependence we have assumed that
$\sigma_0$ has its low temperature value, although near $T_A$, $\sigma$ will
be smaller, leading to faster crossover to the exponential relaxation above $T_A$.

\begin{figure}
\psfig{file=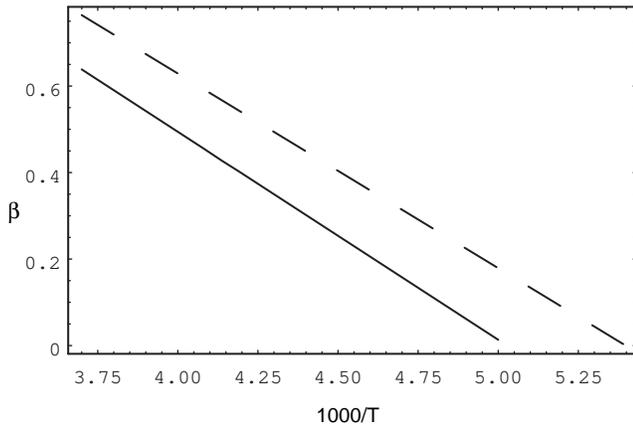,width=3.3in}
\caption
{The temperature dependence of $\beta$ for the most studied fragile glass former, o-terphenyl is shown.  The dashed line contains the experimental measurements of Dixon and Nagel and their extrapolation to lower temperatures [26].  The solid line is
the prediction made with random first order transition theory.  Again no
adjustable parameters are present in the theory.
}
\end{figure}

We see that the mosaic structure expected by the random first order transition approach to the glass transition can quantitatively explains both the trends in
the relaxation dynamics over a range of substances and the temperature dependence of the deviation from Debye behavior. 

This work is supported by NSF grant CHE-9530680.


\begin{thebibliography}{10}

\bibitem{Sillescu99}
H. Sillescu, J. Non-Cryst. Solids {\bf 243},  81  (1999).

\bibitem{KW87a}
T.~R. Kirkpatrick and P.~G. Wolynes, Phys.\ Rev.\ A {\bf 35},  3072  (1987).

\bibitem{KT87}
T.~R. Kirkpatrick and D. Thirumalai, Phys.\ Rev.\ Lett. {\bf 58},  2091
  (1987).

\bibitem{KW87b}
T.~R. Kirkpatrick and P.~G. Wolynes, Phys.\ Rev.\ B {\bf 36},  8552  (1987).

\bibitem{KTW89}
T.~R. Kirkpatrick, D. Thirumalai, and P.~G. Wolynes, Phys.\ Rev.\ A {\bf 40},
  1045  (1989).

\bibitem{MP99}
M. Mezard and G. Parisi, Phys.\ Rev.\ Lett. {\bf 82},  747  (1999).

\bibitem{GKS85}
D.~J. Gross, I. Kanter, and H. Sompolinsky, Phys.\ Rev.\ Lett. {\bf 55},  304
  (1985).

\bibitem{Gardner85}
E. Gardner, Nucl.\ Phys.\ B {\bf 257[FS14]},  747  (1985).

\bibitem{MPV87}
M. Mezard, G. Parisi, and M.~A. Virasoro, {\em Spin Glass Theory and Beyond}
  (World Scientific, Singapore, 1987).

\bibitem{BGS84}
U. Bengtzelius, W. Gotze, and A. Sjolander, J.\ Phys.\ C {\bf 17},  5915
  (1984).

\bibitem{Gotze91}
W. Gotze, {\em Liquids, Freezing and the Glass Transition} (ed. by J. P.
  Hansma, D. Levesque and J. Zinn-Justin, North-Holland, Amsterdam, 1991), pp.\
  287--504.

\bibitem{SW84}
J.~P. Stoessel and P.~G. Wolynes, J.\ Chem.\ Phys. {\bf 80},  4502  (1984).

\bibitem{SSW85}
Y. Singh, J.~P. Stoessel, and P.~G. Wolynes, Phys.\ Rev.\ Lett. {\bf 54},  1059
   (1985).

\bibitem{DV99}
C. Dasgupta and O.~T. Valls, Phys.\ Rev.\ E {\bf 59},  3123  (1999).

\bibitem{Kauz43}
W. Kauzmann, Chem.\ Rev. {\bf 34},  219  (1943).

\bibitem{Wolynes89}
P.~G. Wolynes, {\em Proceedings of the International Symposium on Frontiers in
  Science} (ed. by S. Chan and P. Debrunner, American Institute of Physics,
  1989), pp.\ 39--65.

\bibitem{XW00}
X. Xia and P.~G. Wolynes, Proc.\ Natl.\ Acad.\ Sci.\ U. S. A. {\bf 97},  2990
  (2000).

\bibitem{BNAP93}
B. B\"{o}hmer, K.~L. Ngai, C.~A. Angell, and D.~J. Plazek, J.\ Chem.\ Phys. {\bf
  99},  4201  (1993).

\bibitem{KDPPG97}
W. Kob {\it et~al.}, Phys.\ Rev.\ Lett. {\bf 79},  2827  (1997).

\bibitem{DDKPPG98}
C. Donati {\it et~al.}, Phys.\ Rev.\ Lett. {\bf 80},  2338  (1998).

\bibitem{TWHFSS98}
U. Tracht {\it et~al.}, Phys.\ Rev.\ Lett. {\bf 81},  2727  (1998).

\bibitem{CE95}
M.~T. Cicerone and M.~D. Ediger, J.\ Chem.\ Phys. {\bf 103},  5684  (1995).

\bibitem{AG65}
G. Adam and J.~H. Gibbs, J.\ Chem.\ Phys. {\bf 43},  139  (1965).

\bibitem{Landau69}
G. Parisi, {\em Statistical Physics} (Addison-Wesley, Reading, MA, 1966), pp.\
  348--353.

\bibitem{Ediger98}
M.~D. Ediger, J. Non-Cryst. Solids {\bf 235-237},  10  (1998).

\bibitem{DN88}
P.~K. Dixon and S.~R. Nagel, Phys.\ Rev.\ Lett. {\bf 61},  341  (1988).

\bibitem{CG81}
M.~H. Cohen and G.~S. Grest, Phys.\ Rev.\ B {\bf 24},  4091  (1981).

\end{thebibliography}
\end{document}